\documentstyle[12pt]{article}
\begin{document}
\baselineskip 0.6cm

\newcommand{\gsim}{ \mathop{}_{\textstyle \sim}^{\textstyle >} }
\newcommand{\lsim}{ \mathop{}_{\textstyle \sim}^{\textstyle <} }
\newcommand{\vev}[1]{ \langle {#1} \rangle }

\begin{titlepage}

\begin{flushright}
UCB-PTH-02/11 \\
LBNL-49936 \\
\end{flushright}

\vskip 1.0cm

\begin{center}
{\Large \bf $R$ Symmetry and the $\mu$ Problem}

\vskip 1.0cm

{\large
Lawrence J. Hall, Yasunori Nomura and Aaron Pierce
}

\vskip 0.5cm

{\it Department of Physics, University of California,
                Berkeley, CA 94720, USA}\\
{\it Theoretical Physics Group, Lawrence Berkeley National Laboratory,
                Berkeley, CA 94720, USA}

\vskip 1.0cm

\abstract{A natural origin for the $\mu$ and $\mu B$ parameters of 
weak scale supersymmetric theories is proposed, applicable to
any supersymmetry breaking messenger scale between the weak and Planck 
scales. Although quite general, it requires supersymmetric interactions 
to respect an $R$ symmetry with definite quantum numbers, and it 
requires some new scale of symmetry breaking. The required $R$ symmetry
distinguishes the Higgs boson from the sneutrino, preserves baryon
number in operators of dimension four and five, and contains $R$ parity
so that the lightest superpartner is stable. This origin for $\mu$ 
works for a variety of mediation mechanisms, including gauge mediation, 
gaugino mediation, and boundary condition breaking of supersymmetry.
In any of these mediation schemes, our mechanism leads to a real $B$
parameter, and the supersymmetric $CP$ problem is solved.
This $R$ symmetry may naturally arise from supersymmetric theories 
in higher dimensions.}

\end{center}
\end{titlepage}

{\bf 1.} 
The possibility that nature becomes supersymmetric at the weak 
scale offers a well motivated and exciting scenario for physics beyond
the standard model. It is well motivated because it allows a dynamical
generation of the weak scale, and an understanding of why this scale
is much less than the Planck scale. Furthermore, it gives rise to a
highly successful numerical prediction for gauge coupling
unification. It is exciting because weak scale supersymmetry will be 
thoroughly tested at hadron colliders over the next decade. 

Nevertheless, the underlying structure of the fields and interactions
of the weak scale supersymmetric theory contains three puzzles:
\begin{itemize}
\item  In non-supersymmetric field theories there are three distinct 
 types of fields, corresponding to particles with spin $0$, $1/2$ and $1$. 
 In this case, there is no doubt as to what distinguishes the Higgs field 
 from the lepton doublet field. In contrast, in supersymmetric field 
 theories there are just two types of fields: vector multiplets and 
 chiral multiplets. Hence it is now no longer clear what distinguishes 
 a Higgs field, $H$, from a matter field, $M$. In particular, the down 
 type Higgs and lepton doublets have identical gauge and spacetime 
 properties; what distinguishes the Higgs boson from the sneutrino?
\item  Phenomenologically, interactions cannot be the most general 
 allowed by known gauge and spacetime symmetries. The superpotential must 
 contain interactions of the form $MMH$ for quark and lepton masses, and 
 most probably $MMHH$ at the weak scale for neutrino masses. Yet other 
 forms of superpotential interactions, such as $MH$, $M^3$, and $M^4$ are 
 highly constrained by neutrino masses and proton decay. Some of these 
 interactions are either highly suppressed or forbidden.
\item  A superpotential interaction of the form $HH$ is a special and
 intriguing case. On one hand it must be highly suppressed since a
 coefficient of order the Planck scale or unified mass scale would 
 remove the Higgs doublets from the low energy theory. On the other hand 
 it cannot vanish since otherwise there is a massless charged fermion
 coming from the Higgs/vector multiplets. Indeed the theory is only
 realistic if the coefficient, $\mu$, is of order the weak scale, 
 leading to the well known $\mu$ problem. Why should this supersymmetric 
 parameter be of order the supersymmetry breaking scale?
\end{itemize}

The other great mystery of low energy supersymmetry is the origin of
supersymmetry breaking. Like the supersymmetric interactions, a great
deal about the structure of the supersymmetry breaking interactions is
governed by the requirement of consistency with experiment. However,
nothing determines the ``messenger'' scale, $M_m$: the highest scale at 
which the supersymmetry breaking interactions of squarks, sleptons, 
Higgs and gauginos are local. In supergravity theories this locality is
maintained up to the Planck scale \cite{gravity}, but other methods of 
transporting supersymmetry breaking to the standard model superpartners, 
such as gauge mediation \cite{gaugemed1, gaugemed2}, gaugino mediation 
\cite{gauginomed}, and boundary condition supersymmetry breaking 
\cite{Barbieri:2001yz}, have messenger scales that can be anywhere 
between the weak and Planck scales. 

In this paper we study the consequences of imposing an $R$ symmetry on
the supersymmetric interactions of effective theories with weak scale
supersymmetry.  This symmetry will allow us to address all three of 
the puzzles listed above in the the context of supersymmetry breaking 
with an arbitrary messenger scale.  We will describe a continuous 
symmetry, $U(1)_R$, although a discrete subgroup is sufficient for 
our purposes.  The soft supersymmetry breaking operators break 
the $R$ symmetry, since they include Majorana gaugino masses, 
and therefore holomorphic, $A$ term, scalar interactions.
Using this $R$ symmetry, we find a new mechanism for solving the 
$\mu$ problem for arbitrary messenger scales, and this requires 
a unique choice for the $R$ quantum numbers of matter and Higgs 
fields. Furthermore, we find that this is also the unique choice which 
accounts for the absence of $MH$, $M^3$ and $M^4$ superpotential 
interactions, while allowing consistency with quark-lepton unification. 
Finally, this $R$ symmetry forces a distinction between Higgs and 
lepton doublets.

There are two well-known classes of solutions to the $\mu$ problem.  
One class corresponds to modifying physics at the Planck scale by adding 
non-renormalizable operators \cite{Giudice:1988yz}; the other changes 
the physics at the weak scale, as is the case in the Next-to-Minimal 
Supersymmetric Standard Model (NMSSM) \cite{NMSSM}. The first approach 
requires supersymmetry breaking to be mediated at the Planck scale, while 
the second requires a departure in the weak scale theory from the Minimal 
Supersymmetric Standard Model (MSSM).  Neither of these requirements 
apply to our mechanism. 

In our mechanism, we suppose that all supersymmetric interactions 
respect some global symmetry, $G$, which forbids $\mu$ and commutes 
with both gauge and flavor symmetries.  To allow for the the possibility 
of unification of the matter, we require the $G$ charges of all matter 
multiplets to be equal.  We consider that some field acquires a vacuum 
expectation value (vev) at scale $\Lambda$ between the weak and Planck 
scales.  Supersymmetry breaking interactions, mediated at any scale 
above $\Lambda$, break $G$ and cause a deformation in the vacuum 
resulting in the generation of $\mu$ of order the supersymmetry breaking 
scale.  Such theories have been constructed for the case of mediation 
at the Planck scale \cite{Hempfling:1994ae}. However, for such high 
mediation scales $\mu$ is not a problem since the Giudice-Masiero 
mechanism is available. We study the case of an arbitrary mediation 
scale.  Related solutions for the $\mu$ problem in the context of gauge 
mediation have been proposed \cite{Dvali:1996cu}, but differ from our 
mechanism in the origin of the vacuum deformation.

{\bf 2.}
If we require that our solution to the $\mu$ problem apply to arbitrary 
mediation scales, we find that the operator giving rise to $\mu$ is 
uniquely determined. We assume this operator must be present at tree 
level --- if $\mu$ arises radiatively then the soft $\mu B H_u H_d$ term 
is typically generated at too high a level. Furthermore, $\mu$ must 
arise from a renormalizable operator. If the operator had a coefficient 
suppressed by powers of the Planck scale, it would not be possible 
to get a $\mu$ parameter of the desired size for arbitrary values of 
the messenger scale. Thus the operator generating $\mu$ is unique: 
\begin{equation}
  W_\mu = \lambda X H_{u} H_{d},
\label{eq:XHH}
\end{equation}
where $X$ is a standard model gauge singlet chiral superfield.  Our 
mechanism requires that the supersymmetric interactions break some 
symmetry at a scale $\Lambda$, giving a mass to $X$ and forcing 
$\vev{A_X}=\vev{F_X}=0$.  Here $A_X$ and $F_X$ represent the lowest and 
highest components of the chiral superfield $X$, respectively. Non-zero 
values for $\vev{A_X}$ and $\vev{F_X}$ are generated by supersymmetry 
breaking. 

We discover that our global symmetry, $G$, must be an $R$ symmetry from 
the following argument.  The order of magnitude of $\vev{A_X}$ and 
$\vev{F_X}$ generated after supersymmetry breaking can be understood from 
the $G$ charges of both $X$ and the soft supersymmetry breaking operators.
If $G$ is a non-$R$ symmetry then $A_X$ and $F_X$ have the same $G$ 
transformation, so that both $\mu$ and $\mu B$ are generated at the same 
order in supersymmetry breaking. Hence $B$ is of order $\Lambda$ and 
much too large: our mechanism requires $G$ to be an $R$ symmetry.  

Now, under the $R$ symmetry, there are only two types of supersymmetry 
breaking terms appearing in the scalar potential.  There are holomorphic 
terms, which we denote by $A$ and assign $R$ charge $-2$, and there 
are non-holomorphic terms, denoted $m^{2}$, which have $R$ charge zero.  
A successful solution to the $\mu$ problem requires $\mu$, and 
therefore $\vev{A_X}$, to be linear in supersymmetry breaking. This 
requires that $X$ has $R$ charge $+2$ or $-2$, so that $\vev{A_X}$ 
can be generated proportional to $A^*$ or $A$.  Interestingly, this 
automatically leads to $\vev{F_X}$ of exactly the right order, 
since now $F_X$ has $R$ charge $0$ or $-4$ and is naturally generated 
at second order in supersymmetry breaking proportional to 
$|A|^2, m^2$ or $A^2$. 

Let us consider our two possible $R$-charge assignments separately. 
If $X$ has $R$ charge $-2$, then $H_{u} H_{d}$ has total $R$ charge $+4$.  
Due to the presumed existence of unification, all matter fields carry the 
same $R$ charge.  The Yukawa couplings then force equal $R$ quantum 
numbers for the two Higgs doublets, $R(H_{u})=R(H_{d})$, so that 
$R(H)=2$ and $R(M)=0$.  However, in this case the superpotential 
interaction $MH$ is allowed, which will push some MSSM matter fields 
to have masses of order the Planck scale. Therefore, we prefer to 
consider the charge assignment $R(X)=2$.  Again, the Yukawa couplings 
require $R(H_{u})$=$R(H_{d})$, so we have
\begin{equation}
  R(H) = 0 \;\;\;\;\;\ R(M)=1. 
\label{eq:R}
\end{equation}
This $R$ symmetry is extremely powerful: as well as distinguishing 
between the lepton and Higgs doublet and forbidding $MH$, it also 
forbids $M^3$ and $M^4$, leading to baryon and lepton number 
conservation from operators of dimension four and five.  Therefore, 
our solution to the $\mu$ problem has forced us to forbid dangerous 
dimension four and five operators that might lead to too rapid 
proton decay. Moreover, if the soft supersymmetry breaking operators 
provide the only source of $R$ breaking, then $R$ parity remains 
unbroken, leading to stability of the lightest superpartner.\footnote{
$R$ parity forbids the generation of dimension four baryon and lepton 
number violating operators, even after supersymmetry is broken.
While dimension five proton decay operators could be generated 
with small coefficients proportional to supersymmetry breaking, 
they are phenomenologically irrelevant.}

With a mild assumption about the origin of neutrino masses, a completely 
independent argument will lead us to an identical conclusion for the 
global symmetry $G$. The Yukawa interactions, $M^2H$, possess a 
non-$R$ Peccei-Quinn symmetry (PQ: $M(1), H(-2)$) as well as the $R$ 
symmetry of Eq.~(\ref{eq:R}).  In fact, $G$ must be a linear combination 
of $R$ and PQ, or one of its subgroups. The existence of small neutrino 
masses strongly suggests that the superpotential also contains $M^2H^2$. 
Provided that this interaction is not generated by supersymmetry 
breaking \cite{Arkani-Hamed:2000bq}, this immediately implies that $H$ 
is neutral under $G$, and hence $G$ must be the $R$ symmetry given in
Eq.~(\ref{eq:R}).

{\bf 3.}  The simplest model which realizes the above general mechanism 
for generating the $\mu$ term is given by the superpotential 
\begin{equation}
  W_0 = f X ( Y^2 - \Lambda^2 ),
\label{eq:basic}
\end{equation}
where $R(X) = 2$ and $R(Y) = 0$.  Here we imagine that the scale 
$\Lambda$ is much larger than the weak scale, although this is not 
necessary for our mechanism to work.  We also impose a discrete 
symmetry $Y \rightarrow -Y$, so that the gauge hierarchy is not 
destabilized by the generation of a large tadpole operator for 
a singlet field.\footnote{
We assume that violations of global symmetries by nonperturbative 
gravitational effects \cite{Coleman:1988cy} are sufficiently suppressed.}
Without supersymmetry breaking, the minimum of the potential lies at 
$\vev{A_X} = 0$ and $\vev{A_Y} = \Lambda$, satisfying 
$\vev{F_X} = \vev{F_Y} = 0$.  There is no flat direction 
at this level, and all the fields have masses of order $\Lambda$.  

When we add supersymmetry breaking terms, with a scale $\tilde{m}$
of order the weak scale, the vevs will shift. The most general soft 
supersymmetry breaking terms are given by 
\begin{equation}
  {\cal L}_{\rm soft,0} = - m_X^2 |X|^2 - m_Y^2 |Y|^2 
    - \left( a_f X Y^2 - a_\Lambda \Lambda^2 X + {\rm h.c.} \right).
\end{equation}
Here, $R(m_X^2) = R(m_Y^2) = 0$ and $R(a_f) = R(a_\Lambda) = -2$, and 
we have used $X$ and $Y$ to denote the scalar fields of the respective 
chiral superfields.  By minimizing the scalar potential, we obtain 
$\vev{X} \simeq (a_\Lambda^*-a_f^*)/4|f|^2 \sim \tilde{m}$ and 
$\vev{F_X} \simeq [(a_\Lambda+a_f)(a_\Lambda^*-a_f^*)/4|f|^2 + m_Y^2]/2f 
\sim \tilde{m}^2$.  The vevs of $X$ and $F_X$ are both of the order 
of the weak scale as indicated by the previous general analysis.
Therefore, if we introduce couplings to the Higgs doublet 
$W = W_0 + \lambda X H_{u} H_{d}$ and ${\cal L}_{\rm soft} = 
{\cal L}_{\rm soft,0} - m_{H_{u}}^2 |H_{u}|^2 - m_{H_{d}}^2 |H_{d}|^2 
- (a_\lambda X H_{u} H_{d} + {\rm h.c.})$, $\mu$ and $\mu B$ terms of 
order $\tilde{m}$ and $\tilde{m}^2$ are generated as 
\begin{equation}
  \mu = \lambda \vev{X} 
      \simeq \frac{\lambda(a_\Lambda^*-a_f^*)}{4|f|^2},
\label{eq:mu}
\end{equation}
\begin{equation}
  \mu B = - \lambda \vev{F_X} + a_\lambda \vev{X}
      \simeq \frac{(2f a_\lambda - \lambda a_\Lambda - \lambda a_f)
      (a_\Lambda^*-a_f^*)}{8f|f|^2} - \frac{\lambda m_Y^2}{2f},
\label{eq:muB}
\end{equation}
where $\mu$ and $\mu B$ are defined by $W = \mu H_{u} H_{d}$ and 
${\cal L} = -\mu B H_{u} H_{d} + {\rm h.c.}$.  Although $\mu = 0$ for 
$a_\Lambda = a_f$, it is natural to expect $a_\Lambda \neq a_f$ since 
the two parameters run differently under renormalization 
group evolution; a simple realistic example will be given later.
An important point here is that there exists a parameter region where 
the additional Higgs coupling in Eq.~(\ref{eq:XHH}) does not change the 
gross dynamics at the high scale.  For instance, if $m_{H_{u}}^2$ and 
$m_{H_{d}}^2$ are sufficiently large, the vev of order $\Lambda$ is 
still entirely contained in the $Y$ field, and the vevs for the Higgs 
doublets stay smaller than the weak scale.  Note that here the various 
supersymmetry breaking parameters are evaluated at the scale $\Lambda$. 
Thus, both $m_{H_{u}}^2$ and $m_{H_{d}}^2$ can be positive without 
conflicting with electroweak symmetry breaking.  Below the scale $\Lambda$, 
the heavy fields $X$ and $Y$ are integrated out and the effective theory 
contains only the Higgs doublets with the $\mu$ and $\mu B$ parameters 
given by Eqs.~(\ref{eq:mu}, \ref{eq:muB}).  The soft supersymmetry 
breaking parameters must be further evolved from $\Lambda$ to $\tilde{m}$ 
using the renormalization group equations of this effective theory 
(MSSM), to evaluate electroweak symmetry breaking.

An important requirement for our mechanism is that $\Lambda$ must 
be smaller than the messenger scale of supersymmetry breaking, 
$\Lambda < M_m$.  Therefore, the superpotential Eq.~(\ref{eq:basic}) 
itself is not sufficient for a complete solution of the $\mu$ problem 
(except for the supergravity mediation case), since we have introduced 
by hand a mass parameter, $\Lambda$, smaller than the fundamental 
scale.  A complete solution, however, is obtained if we generate the 
scale $\Lambda$ by the dynamics of strong gauge interactions. Consider, 
for example, the $SU(2)_S$ gauge theory with four doublet chiral 
superfields $Q_i$ ($i=1,\cdots,4$) with the following superpotential:
\begin{equation}
  W_{\rm 0,tree} = f X \left( Y^2 - (QQ) \right) + f' X^a (QQ)_a.
\label{eq:complete}
\end{equation}
This superpotential explicitly breaks a flavor $SU(4)_F$ symmetry of the
$Q_i$ down to $SP(4)_F$; $(QQ)$ and $(QQ)_a$ ($a = 1,\cdots,5$) denote 
singlet and five-dimensional representations of $SP(4)_F$ given by 
suitable combinations of gauge invariants $Q_i Q_j$.  The strong 
dynamics of the $SU(2)_S$ gauge theory is described by the effective 
superpotential
\begin{equation}
  W_{\rm 0,eff} = W_{\rm 0,tree} 
    + S \left( (QQ)^2 + (QQ)_a^2 - \Lambda^4 \right),
\label{eq:strong}
\end{equation}
where $S$ is an additional Lagrange multiplier chiral superfield 
\cite{Seiberg:1994bz}.  For a relatively large value of the coupling 
$f'$, the vacuum lies at $(QQ) = \Lambda^2$ and $(QQ)_a = 0$, so 
that the superpotential $W_{\rm 0,eff}$ is effectively reduced 
to Eq.~(\ref{eq:basic}).  Note that the original tree-level 
superpotential, Eq.~(\ref{eq:complete}), does not contain any mass 
parameters and is invariant under the $U(1)_R$ symmetry with 
$R(X) = R(X^a) = 2$ and $R(Q_i) = 0$.  In fact, it is the most general 
superpotential consistent with the combined $R$ and $SP(4)_F$ 
symmetries. (A linear term in $X$ is forbidden either by requiring 
that the superpotential not contain any mass parameters, or by 
imposing an anomalous discrete $Z_3$ symmetry under which all the 
fields are transformed by $\exp(2\pi i/3)$.)  It is also 
important that $U(1)_R$ does not have an anomaly for $SU(2)_S$ 
(i.e. $\Lambda$ does not carry $U(1)_R$ charge), so that the previous 
general argument is not affected by the strong $SU(2)$ gauge dynamics.

We now consider an application of our mechanism to realistic theories.
We find that the mechanism fits beautifully into the framework where 
small neutrino masses are generated by integrating out right-handed 
neutrino fields through the see-saw mechanism \cite{see-saw}.  
We consider the following theory.  In addition to the usual three 
generations of standard-model quark and lepton superfields, $Q$, $U$, 
$D$, $L$ and $E$, we introduce three right-handed neutrino superfields 
$N$.  Here, we have omitted generation indices.  The Yukawa couplings 
are given by 
\begin{equation}
  W_{\rm Yukawa} = 
    y_u Q U H_{u} + y_d Q D H_{d} + y_e L E H_{d} + y_\nu L N H_{u}.
\label{eq:yukawa}
\end{equation}
We also introduce the $U(1)_X$ gauge symmetry, contained in 
$SO(10)/SU(5)$, under which various fields transform as $Q(1)$, 
$U(1)$, $D(-3)$, $L(-3)$, $E(1)$ and $N(5)$.  This gauge symmetry is 
broken by the vevs of the fields $\Phi(10)$ and $\bar{\Phi}(-10)$ 
through the superpotential
\begin{equation}
  W_{\rm Breaking} 
    = f X \left( \Phi \bar{\Phi} - (QQ) \right) + f' X^a (QQ)_a.
\label{eq:B-L}
\end{equation}
Here $(QQ)$ and $(QQ)_a$ are gauge invariants consisting of $Q_i$, 
the doublets under the strong $SU(2)_S$ gauge interaction (see 
discussion around Eqs.~(\ref{eq:complete}, \ref{eq:strong})).
Note that the above superpotentials, Eqs.(\ref{eq:yukawa}, \ref{eq:B-L}), 
do not contain any mass parameters and are invariant under 
the $U(1)_R$ symmetry, $R(Q) = R(U) = R(D) = R(L) = R(E) = R(N) = 1$, 
$R(H_{u}) = R(H_{d}) = R(\Phi) = R(\bar{\Phi}) = R(Q_i) = 0$ 
and $R(X) = R(X^a) = 2$.  For a relatively large $f'$, the dynamics 
of the $SU(2)_S$ gauge interaction cause the condensation of 
$(QQ) = \Lambda^2$, which is transmitted to the vevs of the $\Phi$ and 
$\bar{\Phi}$ fields, $\vev{\Phi} = \vev{\bar{\Phi}} = \Lambda$.  Here, 
the equality $\vev{\Phi} = \vev{\bar{\Phi}}$ is forced by the $U(1)_X$ 
$D$-term condition.  The vevs for all the other fields are zero at this 
stage: $\vev{X} = \vev{X^a} = \vev{(QQ)_a} = 0$.  After introducing 
soft supersymmetry breaking operators, the vevs of the fields shift.
In particular, non-vanishing vevs for $X$ and $F_X$ are generated as 
$\vev{X} \sim \tilde{m}$ and $\vev{F_X} \sim \tilde{m}^2$, as long as 
holomorphic soft supersymmetry breaking parameters are not subject to 
the special relation $a_{X\Phi\bar{\Phi}} = a_{X(QQ)}$ at the scale 
$\Lambda$. In fact, it is quite natural to expect that the $A$ terms 
for $X \Phi \bar{\Phi}$ and $X (QQ)$ are different since they are 
renormalized differently above the scale $\Lambda$; for example, they 
receive contributions from $U(1)_X$ and $SU(2)_S$ gauginos, 
respectively.  Therefore, by introducing the couplings 
\begin{equation}
  W_{\rm Masses} = \kappa \bar{\Phi} N^2 + \lambda X H_{u} H_{d},
\label{eq:coupling}
\end{equation}
the Majorana masses $M_R$ for the right-handed neutrinos of order 
$M_R \sim \vev{\bar{\Phi}} \sim \Lambda$ and $\mu$ and $\mu B$ 
parameters of order $\mu \sim B \sim \tilde{m}$ are generated.
As in the previous example, the superpotential $W = W_{\rm Yukawa} + 
W_{\rm Breaking} + W_{\rm Masses}$ is the most general 
renormalizable superpotential consistent with the gauge 
$SU(3)_C \times SU(2)_L \times U(1)_Y \times U(1)_X \times SU(2)_S$ 
and global $U(1)_R \times SP(4)_F$ symmetries of the theory, after 
removing a linear term in $X$ as before.

We finally discuss how our general mechanism works explicitly in 
various supersymmetry breaking scenarios.  In gauge mediated 
supersymmetry breaking, our mechanism requires that the mass of the 
messenger fields, $M_m$, is larger than $\Lambda$.  Since the 
holomorphic supersymmetry breaking terms ($A$ terms) required for the 
$\mu$-term generation are small at the messenger scale $M_m$, they must 
be generated by renormalization group evolution from $M_m$ to $\Lambda$.
This can be accomplished, for example, by giving non-trivial $SU(2)_S$ 
or $U(1)_X$ quantum numbers to the messenger fields.  
In the case of gaugino mediation and boundary condition supersymmetry 
breaking, our mechanism requires that the compactification scale is 
larger than $\Lambda$.  In these cases, the relevant $A$ terms 
of order the weak scale may already exist at the compactification scale, 
so we do not necessarily have to rely on renormalization group 
evolution for their generation.\footnote{
In anomaly mediation \cite{Randall:1998uk}, the holomorphic 
supersymmetry breaking parameter associated with the scale 
$\Lambda$ is large --- of the order of the gravitino 
mass $\sim 10~{\rm TeV}$.  Thus $\lambda \sim 10^{-2}$ is needed 
to generate a $\mu$ parameter of the correct size.  Then, to avoid 
a too large $\mu B$ term, a cancellation between two contributions, 
such as $fa_\lambda$ and $\lambda a_\Lambda$ in Eq.~(\ref{eq:muB}), 
is required at the $1\%$ level.}
In any of these mediation mechanisms, $A$ is real in the basis where 
the gaugino masses are real (except for the case of gaugino mediation 
with tree-level $A$ terms), and our origin for $\mu$ and $\mu B$ then 
leads to a real $B$ parameter: the supersymmetric $CP$ problem is solved.

{\bf 4.}  In this paper we have proposed an origin for the parameters 
$\mu$ and $\mu B$ of the minimal supersymmetric standard model, 
which is applicable for any messenger scale, $M_m$.  Although quite 
general, it does require specific symmetries and interactions. Both 
$\mu$ and $\mu B$ parameters arise from the superpotential interaction 
$XH_u H_d$.  A stage of symmetry breaking occurs at some scale 
$\Lambda < M_m$, giving a mass of order $\Lambda$ to $X$, while 
determining $\vev{A_X}=\vev{F_X}= 0$. Providing the form for the 
superpotential is guaranteed by an $R$ symmetry, with the quantum 
numbers of Eq.~(\ref{eq:R}), the soft supersymmetry breaking operators, 
with coefficients $A$ and $m^2$, lead to a small readjustment of the 
vacuum, giving
\begin{equation}
  \mu \approx A^*, \;\;\;\;\; \mu B \approx |A|^2, m^2.
\label{eq:muconc}
\end{equation}
This $R$ symmetry provides a distinction between Higgs and matter
superfields, and forbids superpotential interactions that would
otherwise lead to baryon number violation at too rapid a rate.
Although $R$ is broken by supersymmetry breaking, the discrete $R$ 
parity survives so that the lightest superpartner is stable. In the
case that the original $R$ symmetry is continuous, an $R$ axion will
be produced by the underlying dynamics which breaks supersymmetry.
If all the $R$ breaking effects are generated spontaneously (including 
the constant term in the superpotential needed to cancel the 
cosmological constant), the dominant mass contribution to the $R$ 
axion will come from the QCD anomaly of the $R$ symmetry.  In this 
case, the $R$ axion provides a solution to the strong $CP$ 
problem \cite{Buchmuller:1983ye}.

At first sight our $R$ symmetry appears to be in conflict with 
grand unification: since $R$ forbids $H_u H_d$, it also forbids the
corresponding mass term for the colored Higgs triplets of unified
theories. However, this turns out to be a virtue --- such mass terms
need to be forbidden to avoid too large a proton decay rate mediated
by triplet Higgsino exchange. The colored partners of $H_{u,d}$ must
become heavy by acquiring mass terms coupling them to other colored
states of the theory. This occurs in the missing partner 
\cite{Masiero:1982fe} and Dimopoulos-Wilczek \cite{Dimopoulos:1981xm} 
mechanisms; however, although these mechanisms are consistent with 
an underlying $U(1)_R$ symmetry, in the simplest such models 
$U(1)_R$ is broken at the unification scale, so our $\mu$ generation 
mechanism may not work in these cases. In contrast, in Kaluza-Klein 
grand unification \cite{Hall:2001pg} the desired colored Higgs mass 
terms arise while preserving $U(1)_R$ symmetry, so that our $\mu$ 
generation mechanism works well in this case.

The $U(1)_R$ symmetry is so crucial in providing an understanding of
the form for the interactions in the superpotential, it is important to
seek its origin. Higher dimensional theories are particularly
interesting since they have an enlarged set of supersymmetry
transformations, which results in a global $R$ symmetry in the
equivalent four dimensional description. In the case of a five dimensional
grand unified theory, compactification breaks the unified gauge 
symmetry and also the $SU(2)_R$ symmetry to $U(1)_R$, so that precisely 
the $R$ charges considered here may arise \cite{Hall:2001pg}.

\vspace{5mm}

{\bf Acknowledgements}

Y.N. thanks the Miller Institute for Basic Research in Science 
for financial support.  This work was supported in part by the Director, 
Office of Science, Office of High Energy and Nuclear Physics, of the U.S. 
Department of Energy under Contract DE-AC03-76SF00098, and in part 
by the National Science Foundation under grant PHY-00-98840.

\end{document}